\documentclass[aps,prd,twocolumn,nofootinbib,amsmath,amssymb]{revtex4}
\usepackage{times}
\usepackage{graphicx}

\begin{document}

\title{Spectral Theory of Automorphic Forms and Analysis of Invariant Differential Operators on $SL_3({\bf Z})$ with Applications}
\author{Sultan Catto$^{{\dag}}$, Jonathan Huntley$^a$, Nam-Jong Moh$^a$ and David Tepper$^a$ }
\affiliation{Physics Department\\University Center and The Graduate School \\ The City University of New York \\ 365 Fifth Avenue, New York, NY 10016-4309 \\ and \\ Center for Theoretical Physics\\ The Rockefeller University \\ New York, New York 10021-6399}
\begin{abstract} 
\vskip .10in
We study a variety of problems in the spectral theory of automorphic forms using entirely analytic techniques such as Selberg trace formula, asymptotics of Whittaker functions and behavior of heat kernels. Error terms for Weyl's law and an analog of Selberg's eigenvalue conjecture for $SL_3({\bf Z})$ is given. We prove the following:  
Let $\cal H$ be the homogeneous space associated to the group
$PGL_3(\bf R)$. Let $X = \Gamma{\backslash SL_3({\bf Z}})$ and consider the first non-trivial eigenvalue
$\lambda_1$ of the Laplacian on $L^2(X)$.  Using geometric
considerations, we prove the inequality $\lambda_1 > 3\pi^2/10>2.96088.$
Since the continuous spectrum is represented by the band
$[1,\infty)$, our bound on $\lambda_{1}$ can be viewed as an analogue
of Selberg's eigenvalue conjecture for quotients of the hyperbolic
half space. Brief comment on relevance of automorphic forms to applications in high energy physics is given.  
\vskip .10in
\noindent
\end{abstract}
\vskip 1cm
\maketitle
{\section{Introduction}}
Automorphic forms are one of the central topics of analytic number theory: cornerstone of modern mathematics, they sit at the confluence of analysis, algebra, geometry and number theory with applications in physics of string theories, statistical mechanics, infinite dimensional Lie algebras, cohomology and other areas of mathematical physics, as well as in computer networks. In recent past there has been many important developments in this field. Aside from Langland's programme being the preoccupation of many mathematicians, remarkable progress has been made in the area of Abelian class field theory, structure theory of representation of groups over local fields (such as group representations and the reciprocal relations between automorphic forms and Galois representations), Artin's L-functions and their meromorphy, analytic properties of L-functions of automorphic forms, multiplicity one theorems, complete reducibility of spaces cusp forms, finite dimensionality of spaces of automorphic forms, among others. After being initiated by Poincar\'e, Klein, Gauss, Jacobi, Riemann and Eisenstein, the automorphic forms (particularly modular forms) further flourished through outstanding works of Siegel, Hecke, Maass, Selberg and others. 
\vskip 0.1in

Sophistication of the formalism of string theories, black holes, M-theory, etc. that involves gauge theories, supersymmetry, general relativity, supergravity theories in higher dimensions and quantum mechanics lead to interesting mathematical physics questions and important new insights in pure mathematics. Particularly important are the string dualities that relate different kind of string theories and have remarkable mathematical consequences. Mirror symmetry and algebraic curves in Calabi-Yau manifolds, string duality, where methods from the conformal field theory and from non-perturbative string theory leading to numerous predictions are some examples. Direct relations exist with quantum cohomology, symplectic geometry, Hodge theory, generalized Kac-Moody algebras and theory of automorphic forms (see, for example \cite{1}-\cite{8}). 
In this paper we will concentrate mainly on purely mathematical aspects of the theory, namely the spectral theory of automorphic forms and small and large eigenvalue questions. Some extensions of our formalism to complex Lie groups (in particular to $L$-groups of $GL(4)$ and $Spin(9)$) that makes use of the exceptional group $F_4$, connections to Cayley plane and exceptional Jordan algebra of dimension 27 will be given in another publication \cite{sultan}. For a more in depth connections between automorphic forms and associated physics we refer to Greg Moore's article which also has an extended bibliography {\cite{moore}}.
 
\vskip 0.1in
Eigenfunctions of the Laplace operator on a Riemann manifold are of great interest for physicists working on a variety of problems. Especially the square-integrable eigenstates are of particular importance. Question of their behavior on high energy limits (with respect to the eigenvalues), their concentration onto specific manifolds or sets such as closed geodesics when being on distinguished energy levels, and the distribution law for these levels are some of the problems that need to be worked out in detail.   
\vspace{ 0.4cm}

{\section {Spectral Theory of Automorphic forms, small and large eigenvalue questions and applications}}   

Differential eigenvalue problems on manifolds have played an important role in physics and mathematics for many years. Many important questions in geometry and physics can be phrased in terms of these type of problems. When the manifold is actually a symmetric space one has both an increase in the number of subjects to which these problems are applicable and an increase in the number of techniques that become applicable. For example differential eigenvalue problems on symmetric spaces have applications in representation theory and number theory. Moreover techniques from these subjects become applicable. There is of course a close relationship between the spectral theory of automorphic forms and these subjects. We  study various problems in the spectral theory of automorphic forms and these can all be recast in terms of differential eigenvalue problems on symmetric spaces. The type of problems we investigate can be roughly divided into two types: The first type is the study of the asymptotic distribution of large eigenvalues of the Laplacian on fixed symmetric states where we look for results such as Weyl's law and error terms in this expression. A related problem for large eigenvalues is the multiplicity problem. In dealing with this issue one tries to obtain good upper bounds for the dimension of the space of eigenfunctions of the Laplacian with eigenvalue $\lambda$. Another related problem is the comparison problem, where one attempts to find a finite amount of data that will imply that two automorphic forms (eigenfunctions) are equal. We will seek a solution to this  problem for the case of $SL_3({\bf Z})$. We will also make some attempt on the possibility of extension to other groups. 

\vskip 0.1in
Let us now fix some notation. Let $G$ be a real reductive or semisimple Lie group. We assume that $G$ is compact. It is well known that $G$ has a maximal compact subgroup, that is unique up to a conjugation. We denote this group by $K$. The space $H=G/K$ is a locally symmetric space. Let $\Gamma$ be a discrete subgroup of $G$, and thus of the local isometries of $H$. Let $X=\Gamma \backslash{H}$, and assume that $X$ has finite volume. The space $X$ has a ring of invariant differential operators, one of which may be chosen to be the Laplacian $\Delta$. Let $N(\lambda)$ denote the dimension of the space of eigenfunctions of $\Delta$ on $X$ that are square integrable. If $X$ is compact, we have $N(\lambda)\sim C\lambda^{\frac{n}{2}}$ where $n$ is the dimension of the space. Here $C$ depends only on the dimension $n$. The previous statement is Weyl's law, and is valid for any compact manifold, without any symmetric assumptions. When $X$ is not compact, Weyl's law may well not hold.

\vskip 0.1in
We now specialize. Let $G=SL_3({\bf R})$, and $\Gamma=SL_3({\bf Z})$. Thus $X$ is a rank two symmetric space, and the ring of invariant differential operators has two generators: The Laplacian $\Delta$ and one other that we will denote by $D$. Moreover, $dim~X=5$, so Weyl's law would state $N(\lambda)\sim C\lambda^{\frac{5}{2}}$. A proof of this is given by Stade and Wallace \cite{sw} using the trace formula developed by Wallace \cite{wallace}. Other approaches will also yield this result(see for example S. Miller's Ph.D. thesis (Yale, 1997))\cite{miller}, where he developed a partial trace formula which circumvents some technical difficulties in computing the Selberg trace formula for the quotient $SL_3({\bf Z})\backslash SL_3({\bf R})/SO_3({\bf R})$, and as an application he established the Weyl asymptotic law for the discrete Laplace spectrum, proving that almost all of its cusp forms are tempered at infinity,)but we focus on this as it is a generalization of this method that we plan to pursue. The proof uses test functions chosen so that one can obtain small time asymptotics of the heat kernel. A Tauberian theorem then gives the result. This method, by its very nature, cannot yield an error estimate. It seems that more careful analysis should, however, allow one to obtain an error estimate. By this we mean the following: one may parametrize automorphic forms, which are simultaneous eigenfunctions of the entire ring of invariant differential operators by $r$ parameters where $r$ is the rank of the symmetric space. In the case of products of hyperbolic half planes it was shown that one may give a resonably precise estimate of the number of automorphic forms whose parameters lie in a given disk. It is quite likely that a similar analysis will work for $SL_3({\bf Z})$. Recently, we have obtained the first steps in this program \cite{contemp1},\cite{contemp}. We have shown that $N(\lambda)=C\lambda^{\frac{5}{2}} + O(\lambda^2)$. We expect to be able to improve the result by "localizing it" in the sense described above. To do this a more careful analysis of the smaller terms in the trace formula is required.

\vskip 0.1in

Any Weyl's law with error terms leads to a bound on the multiplicity of eigenfunctions with given Laplace eigenvalue. The eigenfunctions of interest to us are the cusp forms, and a related problem is the following: given two cusp forms $\Phi$ and $\Psi$, with the same representation theory parameters, can one show that if sufficiently many Fourier coefficients agree, that $\Phi=\Psi$. For $\Gamma=SL_2({\bf Z})$ techniques of Maass show that if the Fourier coefficients $a_n, b_n$ agree for $n<C\sqrt{\lambda}$ then the forms are equal. The proof uses the fact that the fundamental domain for $SL_2({\bf Z})$ has only one cusp, and hence in the upper half plane model of hyperbolic geometry the maximum of a cusp will occur at a point with $y>k$, where $k$ is a fixed constant, and knowledge of the asymptotic behavior, with error estimates, of the Whittaker functions that occur in the Fourier expansions of the cusp forms. D. Bump and J. Huntley \cite{huntleybump} had obtained asymptotic expansions for $GL(3,{\bf R})$ Whittaker functions. These are functions of two variables as the rank of the group is two. Let $y_1$ and $y_2$ denote the variables. It is shown that if $y_1 \rightarrow \infty$ and $y_2$ is such that $y_1 y_2^{-\alpha}$ is a positive constant for $\frac{1}{2}<\alpha<\infty$ (the roles of the variables can be reversed), then 

\begin{equation}
W_{\nu_1,\nu_2}\sim y_1^\frac{1}{3} y_2^\frac{1}{3} (y_1^\frac{2}{3} +y_2^\frac{2}{3})^{-\frac{1}{4}} exp(-(y_1^\frac{2}{3} +y_2^\frac{2}{3})^\frac{3}{2})~.
\end{equation}
In this result we suppressed some constants. In the above cited paper a full asymptotic expansion was given, and as one might expect, the higher order terms are not independent of the representation theory parameters. If in addition to the expansion, error estimates can be developed and estimates for $W_{\nu_1,\nu_2}$ can be found for $\alpha\geq -\frac{1}{2}$ then it is likely that an analysis similar to that of Maass can be performed on the Fourier expansion for $SL_3({\bf Z})$ cusp forms to yield a similar type of comparison theorem.

\vskip 0.1in 

So far we have discussed large eigenvalue problems. The open questions considered were all on higher dimensional (in fact higher rank) symmetric spaces. We now consider small eigenvalue questions. We first consider some higher dimensional questions, and then will consider some problems for classical hyperbolic Riemann surfaces. By small eigenvalue we will mean one that does arise from representation theory parameters that are not induced from unitary characters, and are not in the discrete series. These representations are not tempered, meaning that the matrix coefficients fail to be in $L^{2+\epsilon}$ for some $\epsilon >0$. For rank one spaces, small eigenvalues correspond to eigenvalues that are below the continuous spectrum of $H$. In higher rank the relationship is more complicated. One of the basic questions for small eigenvalues is whether or not they actually exist. A related question, the Ramanujan conjecture, asserts that all representations that yield cuspodial eigenforms are tempered. This is known to be false for certain non-arithmetic groups, however for groups such as $GL(n,{\bf R})$, where all discrete co-finite subgroups are known to be congruence groups the conjecture is expected to be true.

\vskip 0.1in

There have been suggestions that a careful analysis of the ring of invariant differential operators will for $SL_3({\bf Z})$ yield a proof of the conjecture. In this case the conjecture amounts to saying that the representation theory parameters should satisfy Re$\nu_1$=Re$\nu_2 =1/3$. It seems likely that there should be eigenvalues that "almost" violate the Ramanujan conjecture. This would amount to having the imaginary parts of the parameters almost equal in absolute value and opposite sign. A study, using the trace formula might shed some light on the situation if sufficiently sharp results can be obtained.

\vskip 0.1in 

It is, however, possible that somewhat weaker but still very interesting results are possible. As mentioned above, the matrix coefficients of the tempered representations are in the $L^{2+\epsilon}$ for all $\epsilon>0$. Let $p(\pi)$ denote the smallest $p$ such that, for a given cusp form, the associated matrix coefficients are in $L^{p+\epsilon}$. It is conjectured that for the congruence group of level $q$, $\Gamma(q)$ is $G(Z)$, the integer lattice of $G$. One can think of the relationship between $SL_n({\bf R})$ and $SL_n({\bf Z})$ that multiplicity $m(\pi,\Gamma(q))$ is $O(\Gamma(q)^{\frac{2}{P(\pi)}+\epsilon})$. It is known that this conjecture is implied when $G$ is rank one, conjectural result that counts lattice points in $\Gamma(q)$. That conjecture is proved for $SL_n({\bf Z})$ in Huntley and Katznelson's paper \cite{katznelson}. The proof that the lattice point counting conjecture implies "$L^p$" conjecture, as the above result is often called, uses techniques from harmonic analysis (essentially a Poisson summation argument) and a study of the eigenfunctions in the cusps of the space. In higher rank, the cusps are more complex and the harmonic analysis is in several variables. The latter problem seems tractable. The analysis of the cusps present several difficulties, however it seems that these can be overcome. 

\vskip 0.1in
We now consider some problems on hyperbolic Riemann surfaces. A fundamental difference here is that discrete finite co-volume subgroup of isometries of the upper half plane are deformable. More precisely, we have surfaces of the same topology with different metric structure. These structures may be parametrized in terms of moduli space. The behavior of small eigenvalues as we approach the boundary of moduli space is very interesting and worthy of detailed study.

\vskip 0.1in
We will now give a solution to the following problem \cite{catto}: Let $\cal H$ be the homogeneous space associated to the group $PGL_3(\bf R)$. Let $X = \Gamma{\backslash H}$ where $\Gamma =
SL_{3}(\bf Z)$ and consider the first non-trivial eigenvalue
$\lambda_1$ of the Laplacian on $L^2(X)$.  Using geometric
considerations, we prove the inequality $\lambda_1 > 3\pi^2/10$.
Since the continuous spectrum is represented by the band
$[1,\infty)$, our bound on $\lambda_{1}$ can be viewed as an analogue
of Selberg's \cite{selberg} eigenvalue conjecture for quotients of the hyperbolic
half space.   
\vspace{0.4cm}
{\section {Statement of the main theorem}}

A fundamental question in the spectral theory of automorphic forms is
whether small eigenvalues exist.  More specifically, let $G$ be a
noncompact reductive group with finite center, $\Gamma$ a nonuniform
lattice, $K$ a maximal compact subgroup of $G$, and set $X =
\Gamma\backslash G/K$.  It is well known from the theory of
Eisenstein series that $L^2(X)$ has continuous spectrum for the
ring of invariant differential operators, and in particular for the
positive Laplacian, $\bf \Delta$.  The continuous spectrum will be,
in cases of interest such as $PGL_n(\bf R)$, an interval $[a,\infty)$
with $a > 0$.  The question we referred to above is:  Do non-constant
square integrable eigenforms exist with eigenvalue $\lambda < a$?
This problem is important for various considerations in number
theory.  In the case $G=PGL_{2}({\bf R})$ and $\Gamma$ is a
congruence subgroup, Selberg conjectured that no such small
eigenvalues exist.  

\vskip .10in
Here we consider the case when $G = PGL_3(\bf R)$ and $\Gamma
= SL_3(\bf Z)$.  Our main result is the following.
\vskip .10in
{\,\,\,\,\,\bf Theorem} {\it
Let $\lambda_1$ denote the eigenvalue for the first nontrivial
eigenform on $L^2(X)$.  Then $\lambda_1 > 3\pi^{2}/10 > 2.96088$.}

\vskip .10in
\noindent
\section {Notation}

Let $\cal H = G/K$ and set $X = \Gamma\backslash \cal H$.  Explicit
coordinates for $\tau \in \cal H$ via the Iwasawa decomposition are given 
by  

\begin{equation}
\tau = 
\left(\begin{array}{ccc}
 1&x_{2} & x_{3}  \\ 0 & 1 & x_1 \\ 0 & 0 &1
\end{array}\right)
\left(\begin{array}{ccc}
 y_1 y_2 & 0 & 0 \\ 0 & y_1 & 0 \\ 0 & 0 &1
\end{array}\right)
\end{equation}
with $y_{1}, y_{2} > 0$, 
from which one can compute that the (positive) Laplacian $\bf \Delta$. We now explain how to do this:

We now want to introduce coordinates on ${\cal{H}}$. To do this, let us first introduce an auxiliary coordinate $x_4$ as:

\begin{equation}
x_1 x_2 = x_3 + x_4
\end{equation}
The reason for doing this is, if we let

\begin{equation}
w_1= \left(\begin{array}{ccc}
  & & -1 \\
 & -1 & \\
-1 & & 
\end{array} \right)
\end{equation}
then $G$ possesses an involution

\begin{equation}
\i:~~~~~~ g\rightarrow w_1 \cdot ^{t}g^{-1}\cdot w_1
\end{equation}
which preserves the Iwasawa decomposition, and hence induces an involution on ${\cal{H}}$. In terms of the coordinates, $\i$ has the effect

\begin{eqnarray}
x_1 \rightarrow -x_2 \\ \nonumber
x_2 \rightarrow -x_1 \\ \nonumber
x_3 \rightarrow x_4 \\ \nonumber
y_1 \rightarrow y_2 \\ \nonumber
y_2 \rightarrow y_1 
\end{eqnarray}
hence the reason for introducing the auxiliary coordinate $x_4$. We note that involution plays an important role in $GL(3)$ theory. 

\vskip 0.1in

Let $W$ be the group of following six matrices that we identify with the Weyl group of $G$:

\begin{equation}
w_0= \left(\begin{array}{ccc}
1  & & \\
 & 1 & \\
 & & 1 
\end{array} \right)
,~~~~~~~~~~~~w_1= \left(\begin{array}{ccc}
  & & -1\\
 & -1 & \\
-1 & & 
\end{array} \right) \nonumber
\end{equation}

\begin{eqnarray}
w_2= \left(\begin{array}{ccc}
  &-1 & \\
 -1 &  & \\
 & & -1 
\end{array} \right)
,~~~~~~w_3= \left(\begin{array}{ccc}
-1  & & \\
 &  & -1 \\
 & -1 & 
\end{array} \right) \nonumber
\end{eqnarray}

\begin{eqnarray}
w_4= \left(\begin{array}{ccc}
  & 1 & \\
 &  & 1 \\
 1 & & 
\end{array} \right)
,~~~~~~~~~~~~w_5= \left(\begin{array}{ccc}
  & & 1 \\
1 &  & \\
 & 1 & 
\end{array} \right) 
\end{eqnarray}

We now introduce a generalized action of $W$ on two complex variables $\nu_1, \nu_2$. Let us consider the function

\begin{equation}
I(\tau)= I_{(\nu_1, \nu_2)}(\tau) = y_1^{2\nu_1+ \nu_2}~y_2^{\nu_1+2\nu_2}
\end{equation}
on ${\cal{H}}$, in terms of the coordinates. The action of $W$ on the parameters $\nu_1, \nu_2$ is defined by requiring

\begin{equation}
I_{(\nu_1-\frac{1}{3}, \nu_2-\frac{1}{3})}(\tau) = I_{(\mu_1-\frac{1}{3}, \mu_2-\frac{1}{3})}(w\cdot \tau)
\end{equation}
when

\begin{equation}
(\mu_1,\mu_2)=w\cdot (\nu_1, \nu_2)  \label{dd}
\end{equation}
if

\begin{equation}
x_1=x_2=x_3=x_4=0.
\end{equation}
Therefore,

\begin{eqnarray}
w_0\cdot (\nu_1, \nu_2) = (\nu_1, \nu_2) \\ \nonumber
w_1\cdot (\nu_1, \nu_2) = (\frac{2}{3}-\nu_2, \frac{2}{3}-\nu_1) \\ \nonumber
w_2\cdot (\nu_1, \nu_2) = (\nu_1+ \nu_2-\frac{1}{3}, \frac{2}{3} -\nu_1) \\ \nonumber
w_3\cdot (\nu_1, \nu_2) = (\frac{2}{3}-\nu_1, \nu_1+\nu_2-\frac{1}{3}) \\  \nonumber
w_4\cdot (\nu_1, \nu_2) = (1-\nu_1- \nu_2, \nu_1)  \\ \nonumber
w_5\cdot (\nu_1, \nu_2) = (\nu_2, 1-\nu_1 -\nu_2)
\end{eqnarray}

In addition to the parameters $\nu_1, \nu_2$, it is convenient to introduce three equivalent parameters $\alpha$, $\beta$ and $\gamma$ defined by

\begin{eqnarray}
\alpha= -\nu_1 -2\nu_2 +1 \\ \nonumber
\beta= -\nu_1+\nu_2 \\ \nonumber
\gamma= 2\nu_1 +\nu_2 -1
\end{eqnarray}
so that

\begin{equation}
\alpha+\beta+\gamma=0~.
\end{equation}

Clearly, the action (\ref{dd}) of the Weyl group permutes these three quantities. 

We now consider $G$-invariant differential operators on ${\cal{H}}$ that form a commutative algebra, isomorphic to a polynomial ring in two variables. Two particular generators $\bf \Delta$ and $\bf \Delta_3$ can be exhibited:

\begin{equation}
{\bf{\Delta}} I = \lambda I
\end{equation}

\begin{equation}
{\bf{\Delta_3}} I =\mu I
\end{equation}
where 

\begin{equation}
\lambda = -1-\beta \gamma - \gamma \alpha -\alpha \beta
\end{equation}
and

\begin{equation}
\mu= -\alpha \beta \gamma~.
\end{equation}

At this point we can follow procedure given in chapter 2 of {\cite{Bu 84}).  We arrive (after some algebra) at the (positive) Laplacian $\bf \Delta$ which can be written as

\begin{eqnarray}
-{\bf \Delta} &=& y^2_1 \frac{\partial^2}{\partial y^{2}_{1}} -
y_1y_2\frac{\partial^2}{\partial y_{1}\partial y_{2}} + y^2_2
\frac{\partial^2}{\partial y_{2}^{2}} 
+ y^2_1(x^2_2 + y^2_2) \frac{\partial^{2}}{\partial x^{2}_{3}} \nonumber \\ & &+
y^2_1 \frac{\partial^{2}}{\partial x_{1}^{2}} + y^2_2
\frac{\partial^{2}}{\partial x^{2}_{2}}
+ 2y^2_1 x_2 \frac{\partial^{2}}{\partial x_{1}\partial
x_{3}}
\end{eqnarray}
The ring of invariant differential operators is spanned by the identity, the Laplacian
$\bf \Delta$, and a third order operator $\bf \Delta_3$.  The invariant volume element is given 
by 

\begin{equation}
dV =\frac{dx_1dx_2dx_3dy_1dy_2}{y^{3}_{1}y^{3}_{2}}.
\end{equation}
We shall not use the explicit formula for the invariant volume
element; however, the above expression for the Laplacian will be
necessary in our proof of the main theorem.

\vskip .10in
For our purposes, it is more convenient to work with functions on
${\cal H}$ that are $SL_{3}({\bf Z})$ invariant rather than considering
functions on the quotient space $X$.  To this end, we introduce a
fundamental domain for $\Gamma\backslash \cal H$. Specifically,
computations on page 56 of {\cite{Gr 93}} show that a fundamental
domain $\cal D$ is described through the following set of
inequalities: 

\begin{equation}
\qquad \qquad
\begin{tabular}{l}
$v^{\frac{3}{2}} < v^{\frac{3}{2}} (1-x_2 + x_3)^2 + w(1-x_1)^2 +w^{-1};$   \\
$v^{\frac{3}{2}} < v^{\frac{3}{2}}(x_2-x_3)^2 + w(1-x_1)^2 + w^{-1}_1;$ \\
$v^{\frac{3}{2}} < v^{\frac{3}{2}} x^2_2 + w; 
\,\,\,\,\,\,\,
v^{\frac{3}{2}} < v^{\frac{3}{2}} x^2_3 + wx^2_1 + w^{-1};$ \\
$1 < w^{-2} + x^2_1; \,\,\,\,\,\,\,
0 < x_1 < \frac{1}{2} $;\\ 
$ 0 < x_2 < \frac{1}{2} \,\,\,\,\,\,\,
-\frac{1}{2} < x_3 < \frac{1}{2},$ 
\end{tabular} 
\end{equation}
where we have used the notation $w^{-1} = y_1$ and $v^{-\frac{3}{2}}
= y^2_2 y_1$.  Let $S$ denote the Siegel set described via the
inequalities

\begin{eqnarray} 
0 < x_1 < \frac{1}{2}, \,\,\,\,\,
0 < x_2 < \frac{1}{2}, \,\,\,\,\,
-\frac{1}{2} < x_3 < \frac{1}{2}, \nonumber \\ 
y_1 > \frac{\sqrt{3}}{2},\,\,\,\,\, 
y_2 >\frac{\sqrt{3}}{2}.\,\,\,\,\,\,\,\,\,\,\,\,\,\,\,\,\,\, 
\end{eqnarray}
The set $S$ contains the fundamental domain $\cal D$.  Further,
results from page 61 {\cite{Gr 93}} show the existence of elements
$\gamma_{1}, ..., \gamma_{10} \in SL_{3}({\bf Z})$ such that
$S \subset \bigcup\limits^{10}_{i=1} {\cal D} {\gamma_{i}}$ (we have used
the notation $\cal D \gamma$ to denote the image of the fundamental
domain $\cal D$ under left multiplication by $\gamma$).
The main aspects of the above points which we shall use are the
assertions that for any $\tau \in S$ we have $y_{1}(\tau) >
\sqrt{3}/2$ and that $S$ is contained in ten translates of $\cal D$.

\vskip .10in
Recall that an automorphic form is a $C^{\infty}$ function $\phi$ on
$\cal H$ which satisfies the following properties:
\begin{quote}
(i)~ $\phi (\gamma\circ\tau) = \phi(\tau)$ for $\gamma\in
SL_3(\bf Z)$;

(ii)~ $\vert\phi (\tau)\vert \ll y^{N_{1}}_1 y^{N_{2}}_2$
for $\tau\in \cal D$ and integers $N_1, N_2$;

(iii)~ $\phi$ is an eigenform for the ring of invariant
differential operators.
\end{quote}
\noindent
An automorphic form is said to be a cusp form if it satisfies the
additional property
\begin{quote}
(iv)~

\begin{eqnarray}
& & \int\limits^{\frac{1}{2}}_{-\frac{1}{2}}
\int\limits^{\frac{1}{2}}_{-\frac{1}{2}} \phi\left(
\left[\begin{array}{ccc}
 1 & 0 &\xi_3 \\ 0 & 1 & \xi_1\\ 0 & 0 & 1
\end{array}\right]
 \tau\right) d\xi_1
d\xi_3=  \nonumber \\
&  & \,\,\,\, =\int\limits^{\frac{1}{2}}_{-\frac{1}{2}}
\int\limits^{\frac{1}{2}}_{-\frac{1}{2}}  \phi\left(
\left[\begin{array}{ccc}
 1 & \xi_2 & \xi_3 \\ 0 & 1 & 0 \\ 0 & 0 & 1
\end{array}\right]
\tau\right) d\xi_2
d\xi_3 = 0 
\end{eqnarray}
\end{quote}
\noindent
Cusp forms are square integrable.  Although we shall not need this
fact, let us note that, from the theory of Eisenstein series, the
only noncuspidal square integrable automorphic forms on $X$ are
constant.   

\vskip .10in
\noindent
{\section {Proof of the main theorem}}
\flushleft
Our method of proof is a modification of that used by Roelcke \cite{roelcke} to
show that the small eigenvalue $\lambda_{1}$ for the quotient space $SL_2({\bf Z})\backslash SL_2({\bf R})/ SO_2({\bf R})$
satisfies the bound $\lambda_1 > 3 \pi^{2}/2$ (see page 511 of
{\cite{He 83}}).  We shall use the Fourier expansion of automorphic
forms associated to $SL_{3}(\bf Z)$, as developed in Chapter IV of
{\cite{Bu 84}}.  

\vskip .10in
Assume that $\phi$ is a non-constant automorphic form, so then
${\bf \Delta}\phi =\lambda\phi$ and $\phi{\bf \Delta}\phi =
\lambda\phi^2$.   Through integration by parts, using the automorphic
boundary conditions, and the fact that the Siegel domain $S$ is
contained in ten translates of the fundamental domain $\cal D$, we
obtain the inequality  

\begin{equation}
\frac{\int\limits_S\vert\nabla\phi\vert^{2} dV}
{\int\limits_S \vert\phi\vert^{2} dV}  < 10 \lambda.
\end{equation}
As on page 67 of {\cite{Bu 84}}, let us expand $\phi$ in a Fourier
expansion with respect to the abelian group

\begin{equation}
\left\{
\left[
\begin{array}{ccc}
 1 & 0 & \xi_3 \\ 0 & 1 & \xi_1 \\ 0 & 0 &1
\end{array}
\right]
 \,\,\,\,\,\xi_{1}, \xi_{3} \in \bf R \right\}.
\end{equation}
Specifically, we have $\displaystyle\phi (\tau ) =
\sum\limits_{n_{1},n_{3}}\phi^{n_{3}}_{n_{1}} (\tau)$ 
where

\begin{equation}
\phi^{n_{3}}_{n_{1}}(\tau ) = \int\limits^1_0 \int\limits^1_0
\phi\left(
\left[
\begin{array}{ccc} 
1 & 0 & \xi_3\\ 0 & 1 & \xi_1 \\ 0 & 0 & 1
\end{array}
\right)
\tau\right] e^{-2\pi i(n_{1}\xi_{1}+n_{3}\xi_{3})} d\xi_1
d\xi_3.
\end{equation}
Observe that $\phi^{0}_{0}=0$ since $\phi$ is not constant and square
integrable, hence cuspidal.  Let

\begin{equation}
\Gamma^{2}_{1} = \left\{
\left[
\begin{array}{ccc}
 a & b & 0 \\ c & d & 0 \\ 0 & 0 &1
\end{array}
\right]
 : 
\left[
\begin{array}{cc}
 a & b \\ c & d
\end{array}
\right]
 \in
SL_2({\bf Z})\right\},
\end{equation}
and set $\Gamma^2_\infty$ to be the subgroup of $\Gamma^{2}$ which
stablizes infinity.  As on page 69 of {\cite{Bu 84}}, we then can write
$\phi^{n_{3}}_{n_{1}}$ as 

\begin{equation}
\phi^{n_{3}}_{n_{1}} (\tau) =
\sum\limits_{\gamma\in\Gamma^{2}_{\infty}\backslash\Gamma^{2}_{1}}
\sum\limits_{n_{1}=1}^{\infty}\phi^0_{n_{1}}(\gamma\circ\tau).
\end{equation}
By a standard application of elliptic regularity, $\phi$ is
necessarily $C^\infty$, hence we can interchange integration and
summation and apply Parseval's theorem to obtain the inequality

\begin{equation}
\frac{\int\limits_{S}
\sum\limits_{n_{1}=1}^{\infty}\left|
\sum\limits_{\gamma\in\Gamma^{2}_{\infty}\backslash\Gamma^{2}_{1}}
\nabla_{\tau}\phi^0_{n_{1}}
(\gamma\circ\tau)\right|^2 dV}
{\int\limits_{S}
\sum\limits_{n_{1}=1}^{\infty}
\left|
\sum\limits_{\gamma\in\Gamma^{2}_{\infty}\backslash\Gamma^{2}_{1}}
\phi^0_{n_{1}}(\gamma\circ\tau)\right|^2 dV} < 10 \lambda.
\end{equation}
Since $\nabla$ is an invariant operator, we may differentiate the
expressions in the numerator and then evaluate at $\gamma\circ\tau$,
thus yielding

\begin{equation}
\frac{\int\limits_{S}
\sum\limits^\infty_{n_{1}=1}\left|
\sum\limits_{\gamma\in\Gamma^{2}_{\infty}\backslash\Gamma^{2}_{1}}
\nabla_{\tau}\phi^0_{n_{1}}(\tau)\big|_
{\gamma\circ\tau}\right|^2 dV}
{\int\limits_S\sum\limits^\infty_{n=1}
\left|\sum\limits_{\gamma\in\Gamma^{2}_{\infty}\backslash\Gamma^{2}_{1}}
\phi^0_{n_{1}}(\tau)\big|_{\gamma\circ\tau}\right|^2 dV} < 10 \lambda.
\end{equation}

We now integrate by parts and consider the action of the Laplacian
$\bf \Delta$ on functions of the form $\phi_{n_{1}}^{0}$.  Since each
function $\phi_{n_{1}}^{0}$ is independent of $x_{3}$, these terms in
$\bf \Delta$ annihilate $\phi_{n_{1}}^{0}$.  Observe that all terms
involving $y_{1}, y_{2}$ and $x_{2}$ are positive operators (compare
with line (2.31) on page 32 of {\cite{Bu 84}}), so we obtain the bound

\begin{equation}
{\bf \Delta} \phi_{n_{1}}^{0} \geq - y_{1}^{2} \cdot
\frac{\partial^{2}}{\partial x_{1}^{2}}\phi_{n_{1}}^{0} 
= y_{1}^{2} \cdot 4 \pi ^{2}n_{1}^{2}\phi_{n_{1}}^{0}.
\end{equation}
Since $y^2_1 > \frac{3}{4}$, we have
${\bf \Delta} \phi_{n_{1}}^{0} \geq \frac{3}{4}\cdot 
4 \pi ^{2}n_{1}^{2}\phi_{n_{1}}^{0}=
3 \pi ^{2}n_{1}^{2}\phi_{n_{1}}^{0}$.  Combining this inequality with
the above calculations the cuspidality condition $\phi_{0}^{0} = 0$,
we obtain 

\begin{equation}
10 \lambda >  
\frac{ \int\limits_{S}
\sum\limits^\infty_{n_{1}=1}
\sum\limits_{\gamma\in\Gamma^{2}_{\infty}\backslash\Gamma^{2}_{1}}
{\bf \Delta}_{\tau}\phi^0_{n_{1}}(\tau)\big|_{\gamma\circ\tau} 
\cdot \phi^0_{n_{1}}(\tau)\big|_{\gamma\circ\tau} dV}
{\int\limits_S\sum\limits^\infty_{n=1}
\left|\sum\limits_{\gamma\in\Gamma^{2}_{\infty}\backslash\Gamma^{2}_{1}}
\phi^0_{n_{1}}(\tau)\big|_{\gamma\circ\tau}\right|^2 dV} \geq 3\pi^{2},
\end{equation}
hence $\lambda\geq 3\pi^2/10$.  Since $\phi$ was any cusp form, we
obtain the bound as asserted in the theorem.

\vskip 0.1in
{\section {Concluding remarks}}
\noindent

As the continuous spectrum in this situation is $[1,\infty)$, our
theorem implies an analogue of Selberg's eigenvalue conjecture.
Note that our bound is stronger than result for $SL_{3}(\bf Z)$
from {\cite{Mi 96}} who proved $\lambda_{1} \geq 1$.  In general, our
method applies to $G = SL_{n}(\bf R)$ with $\Gamma = SL_{n}(\bf Z)$ to give the bound $\lambda_{1} > 3\pi^{2}/M$ where $M$ is the
number of fundamental domains which intersect a Siegel set containing
the fundamental domain constructed in {\cite{Gr 93}}; however, for $n
\geq 4$, this bound is rather weak.  Finally, let us remark that our
theorem is indeed a consequence of the Ramanujan conjecture, which
asserts that all nontrivial automorphic representations come from
tempered representations.
\vskip .10in    
({\dag}) Work supported in part by DOE contracts No. DE-AC-0276-ER 03074 and 03075, and PSC-CUNY Research Awards.\\
(a) Also member of Mathematics Department, CUNY; Work supported by the PCS-CUNY Research Awards.

\vfill

}
\enddocument